\documentclass[aps,eqsecnum,showpacs,preprint]{revtex4}

\usepackage{graphicx}
\usepackage{amsmath}
\usepackage{amsfonts}
\usepackage{amssymb}
\usepackage{bm}
\usepackage{psfrag}

\newcommand{\be}{\begin{eqnarray}}
\newcommand{\ee}{\end{eqnarray}}

\begin{document}

\title{Measurement of finite frequency cross-correlations  noise with a resonant circuit}
\author{M. Creux}
\affiliation{ Centre de Physique Th\'eorique, Case 907 Luminy, 13288 Marseille 
cedex 9, France}
\affiliation{Universit\'e de la M\'editerran\'ee, 13288 Marseille cedex 9, 
France}
\author{A. Cr\'epieux}
\affiliation{ Centre de Physique Th\'eorique, Case 907 Luminy, 13288 Marseille 
cedex 9, France}
\affiliation{Universit\'e de la M\'editerran\'ee, 13288 Marseille cedex 9, 
France}
\author{T. Martin}
\affiliation{ Centre de Physique Th\'eorique, Case 907 Luminy, 13288 Marseille 
cedex 9, France}
\affiliation{Universit\'e de la M\'editerran\'ee, 13288 Marseille cedex 9, 
France}

\date{\today}

\begin{abstract}
The measurement of finite frequency cross-correlations  noise represents
an experimental challenge in mesoscopic physics. Here we propose a
generalisation of the resonant
LC circuit setup of Lesovik and Loosen which allow to probe directly such cross-correlations
by measuring the charge fluctuations on the plates of a capacitor. The measuring
circuit collects noise contributions at the resonant frequency of the LC circuit.
Auto-correlation noise can be
canceled out by switching the wires and making two distinct measurements.
The measured cross-correlations then depend on four non-symmetrized correlators.
This detection method is applied to a normal metal three terminal device.
We subsequently discuss to what extent the measurement circuit can detect
electron-antibunching and what singularities appear in the spectral
density of cross-correlations noise.      
\end{abstract}


\maketitle
\section{Introduction}

The measurement of finite frequency noise in mesoscopic systems
provides a useful diagnosis of quantum transport:
it allows to characterize the carriers which are involved. 
In normal metal conductors\cite{yang,blanter_buttiker}, 
at zero temperature (and in the absence of spatial averaging effects \cite{grabert,levinson}), 
finite frequency correlations exhibit a singularity 
at $\omega= eV$. In normal metal-superconductor 
junctions \cite{torres_martin_lesovik}, Andreev reflection gives rise to 
a singularity at $2eV$ signaling that Cooper pairs enter the superconductor,
in the fractional quantum Hall effect, the tunneling of quasiparticles
in the vicinity of a point contact
leads to a singularity at $e\nu V$, with $\nu$ the 
filling factor\cite{chamon_freed96}.
At the same time, ``zero'' frequency noise cross correlations can be used 
to probe directly the statistics of the charge carriers: are the carriers
bunched or anti-bunched\cite{martin_landauer}? This implies a ``Y'' 
geometry where carriers are injected in one arm, and correlations are 
measured in the two receiving arms. 
The purpose of the present work is to discuss an inductive coupling
setup which can be used toward the measurement of high frequency noise correlations
in such geometries. 

Any setup for measuring noise involves the selection (filtering) of a range
of specific  frequencies by the electronic detection circuit. 
Provided that the electrical apparatus in an experiment
can sample the current at discrete times,  
finite frequency noise can in principle 
be computed directly from this time series.
Yet, a noise correlator is a quantum mechanical average of products 
of operators. Symmetrized or non-symmetrized correlators 
can thus be constructed from the time series, 
provided that the frequency is much smaller than
the inverse time step of the series. 
This recipe for computing the noise is not practical at high frequencies.
Here, we are interested in
systems where noise is detected via 
a measuring device -- coupled to the mesoscopic device which should 
pick up the noise contribution at a specific frequency, 
via repeated measurements. 
Such proposals have been put forth and some have been implemented experimentally
within the last decade. Ref. \cite{schoelkopf} uses effectively a LC circuit
in order to measure the noise of a two terminal diffusive conductor.
A theoretical suggestion due to Lesovik and Loosen\cite{lesovik_loosen} consists of a LC circuit 
coupled inductively to the fluctuating current emanating from 
a mesoscopic conductor: the measurement of the charge fluctuations on the 
capacitor plates provides information on the finite frequency noise correlators
at the resonant frequency of the LC circuit. The fact that this method samples 
contributions from the non-symmetrized noise correlators, which are related to 
emission and absorption from the device, has been emphasized in Ref. \cite{gavish}.  
Aguado and Kouvenhoven\cite{aguado_kouwenhoven} have proposed to measure the noise of an arbitrary
circuit by coupling capacitively this circuit to a detector circuit:
a DC current -- generated by inelastic electron tunneling events -- flows 
in the detector circuit when a ``photon''
$\hbar\omega$, is
provided/absorbed by the mesoscopic circuit.  
This theoretical suggestion has been successfully implemented to measure the 
finite frequency noise of a Josephson junction using a large 
Superconductor-Insulator-Superconductor junction\cite{deblock_science}:
quasiparticles tunnelling in the SIS junction can occur only  if it is assisted 
by the frequency  provided by the antenna. A subsequent experiment allowed to isolate 
the emission and absorption contributions to noise \cite{deblock2}.
High frequency noise measurements have been also performed  
using the detection of photons 
emitted by the mesoscopic conductor \cite{gabelli}. 
Such photons propagate in coaxial lines and are analyzed/detected in a 
Hanbury--Brown and Twiss geometry for microwave photons 
appeared in order to measure two terminal noise with an auxiliary 
mesoscopic circuit. 
Turning to noise cross correlations, 
low frequency noise measurement in such branched circuits 
have been performed\cite{henny_oliver}, 
which showed that a fully degenerate electron gas has negative noise 
correlations.  Cross-correlations noise have obvious applications 
in the study of electron transport in
Hanbury--Brown and Twiss type geometries. 
Below, we will mention two situations where they are useful:
the study of electronic entanglement in mesoscopic devices
\cite{lesovik_martin_blatter01,recher_sukhorukov_loss}, and 
the identification of anomalous charges in Luttinger 
liquid wires\cite{lebedev_crepieux_martin}.
Here we consider the case of a setup where the cross-correlations  noise
are analyzed via inductive coupling to the mesoscopic circuit to be measured. 

\begin{figure}
\center{\includegraphics[width=9cm]{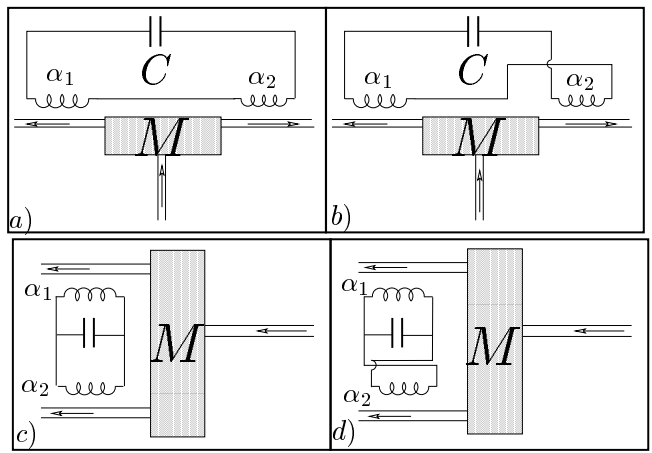}}
\caption{Schematic description of the noise cross-correlation setup. $M$ is 
the mesoscopic circuit to be measured, $C$ is the capacitor and there are 
two inductors with coupling constants $\alpha_1$ and $\alpha_2$
to the mesoscopic circuit. 
$a)$ and $b)$: the electrical components of the detector are in series 
and they ``see'' the current with the same sign ($a$) or 
with the opposite sign ($b$). $c)$ and $d)$: the electrical 
components of the detector are in parallels and they ``see'' the current 
with the same sign ($c$) or with the opposite
sign ($d$).}
\label{fig:setup}
\end{figure}

The paper is organized as follows. In section II, the model with a measuring circuit is
introduced, and it is shown how cross-correlations can be directly 
obtained. In section III, we compute the measured cross-correlations 
noise. In section IV, we introduce a model with two separated circuits
and we compute the measured cross-correlations noise.
 Our results, in the case of one circuit, are then applied in section V 
to a simple device: a three terminals normal metallic sample, where we stress
the difference between the symmetrized noise, the non-symmetrized noise, 
and the measured noise. Section VI gives two examples where cross-correlations
are needed. We conclude in section VII.

\section{Model and method}

For measuring cross-correlations, two inductances 
($L_1$ and $L_2$) and a single capacitor ($C$)
are needed. 
The two inductors having coupling constants
 $\alpha_1$ and $\alpha_2$, 
are placed next to the two outgoing arms
of the three terminal mesoscopic device
(Fig.~\ref{fig:setup}). We consider two cases:
the two inductances are placed in series
(Fig.~\ref{fig:setup}a and Fig.~\ref{fig:setup}b) 
or in parallel 
(Fig.~\ref{fig:setup}c and Fig.~\ref{fig:setup}d). 
Depending on the wiring of these inductances
(Fig.~\ref{fig:setup}a,c or Fig.~\ref{fig:setup}b,d), 
the two inductances ``see'' the outgoing 
currents with the opposite sign or with 
the same sign. 
Throughout this paper, we neglect the mutual inductance of 
the detector circuit: it is justified by the fact that 
such inductances are placed near  opposite ends of the  
circuit whose noise correlations are measured.  
Classically, the charge $x$ on the capacitor plates
of the measuring circuit obeys 
the equation of motion:
  \be
    M\ddot{x}(t)=-D x(t)-\alpha_1\dot{I}_1(t)\mp\alpha_2\dot{I}_2(t)~,
  \ee
where the ``mass'' $M=L_1+L_2$ 
if the circuit is in series (Fig.~\ref{fig:setup}a,b) 
or $M=L_1L_2/(L_1+L_2)$ 
if the circuit is in parallel (Fig.~\ref{fig:setup}c,d), 
and $D=1/C$. The characteristic frequency of the $LC$ circuit is $\Omega=\sqrt{D/M}$.

The currents which appear in the two coupling 
terms are $I_{1(2)}(t)=l^{-1}\int_{x_{1(2)}-l/2}^{x_{1(2)}
+l/2}I(r,t)dr$, where $l$ is the length of 
the inductive coupling region. We assume in a standard way that
the chemical potential of the leads is the largest energy scale, 
compared to the bias voltage and the frequency we want to probe.  
Spatial averaging effects for the currents
have been discussed in Ref. \cite{gavish_prl,gavish_phd}:
the current operator contains fast and slow oscillations
\cite{martin_houches}, at least in the ballistic case.
Here we claim to be measuring a current which is spatially independent 
because of the integral over the inductive region $l$.
We first assume that $l\gg\lambda_F$, so that short wave lengths
oscillations are averaged out. Furthermore, the slowly oscillating 
terms in the current will be reduced to a constant contribution 
if we require that $l\ll v_F/\omega$.
Ref. \cite{grabert} has studied the 
spatial oscillations of the current in the presence of Coulomb interactions: 
interactions modify the wavelength $\pi v_F/\omega$ of such oscillations, 
but the amplitude of its signal is reduced with 
increasing interaction strength.

The derivation of Ref.~\onlinecite{lesovik_loosen} is generalized to the 
two inductances situation. 
To quantize the measuring circuit, we note that the equation 
of motion can be derived from the Hamiltonian: 
 \be
    H=H_0+H_{int}=\frac{p^2}{2M}+\frac{Dx^2}{2}
      +H_{int}~,
  \ee
with 
\be
H_{int}=-\frac{p(\alpha_1I_1(t)\pm\alpha_2I_2(t))}{M}+\frac{(\alpha_1I_1(t)\pm\alpha_2I_2(t))^2}{2M}~.
\ee
The mesoscopic circuit plus measuring circuit are assumed to be 
decoupled at $t= -\infty$. The coupling between the two is switched 
on adiabatically, and one can monitor the charge of the capacitor, and its fluctuations,  
in the presence of the fluctuating currents
at time $t=0$: it is a stationary measurement.

The n-th power of the capacitor charge
reads, in the interaction representation: 
 \be 
\langle x^n(0)\rangle=Tr[e^{-\beta H_0}U^{-1}(0)x^n(0)U(0)]~,
 \ee
 with the evolution operator
 \be
    U(0)=T\exp{\Big(\frac{-i}{\hbar}\int_{-\infty}^0dt'H_{int}(t')\Big)}~.
  \ee  
We calculate perturbatively the $n$-th power of the charge,  
expanding the evolution term ($U(0)$) in powers of $H_{int}$.
Considering the average charge and the average charge square, one obtains:
\be
\langle x(0)\rangle&=&
\langle x(0)\rangle_1+
\langle x(0)\rangle_3+...\ ,
\\
\langle x^2(0)\rangle&=&
\langle x^2(0)\rangle_0+
\langle x^2(0)\rangle_2+...\ ,
\ee
where the different orders in $H_{int}$
are linked to the higher cumulants of the current:
$\langle x(0)\rangle_1$ contains information 
about the average current, $\langle x^2(0)\rangle_2$
about the current fluctuations, 
and $\langle x(0)\rangle_3$ about the third moment, 
and so on. 

The zero order contribution of the charge fluctuations gives: 
 \be
\langle x^2(0)\rangle_0=\frac{\hbar}{2M\Omega}(N(\Omega)+1/2)~,
  \ee
with  $N(\Omega)=1/(e^{\beta\hbar\Omega}+1)$ is the Bose Einstein distribution 
at the detector circuit temperature, which is not necessarily the 
mesoscopic device temperature. The next non-vanishing term, which 
depends on products of current operators is:
 \be
    \langle x^2(0) \rangle_{2\pm}=&&
    \Big\langle
              \frac{1}{2M}\Big(\frac{-i}{\hbar}\Big)^2
                \int_{-\infty}^{0}dt_1
                \int_{-\infty}^{t_1}dt_2        \nonumber\\
                 &\times    & \left[
                        [x(0), p(t_1)(\alpha_1I_1(t_1)
                          \pm\alpha_2I_2(t_1))],\right.\nonumber\\
                &&\left.        p(t_2)(\alpha_1 I_1(t_2)
                        \pm\alpha_2 I_2(t_2))\right]
                     \Big\rangle\nonumber\\
		&- & \frac{1}{2M}\Big(\frac{-i}{\hbar}\Big)
                \int_{-\infty}^{0}dt \langle I^2(t)\rangle\langle x^2(0)\rangle -\langle x^2(0)\rangle \langle I^2(t)\rangle~,
  \label{two_commutators}
  \ee
where one recalls that the sign in front of the coupling constants 
$\alpha_{1,2}$ reflect the choice of the circuit, (a or b) and (c or d).
The calculation of the charge fluctuations gives four terms: two 
autocorrelation terms which correspond to the fluctuations due to a single 
inductor (in terms of $\alpha_1^2$ or 
$\alpha_2^2$), and two terms associated with the correlation 
between the two inductors, which are proportional to $\alpha_1\alpha_2$.
It is precisely these latter terms which allow to detect the noise correlations. 
Thus, it seems that it is impossible in practice to get rid 
of the autocorrelation terms: the measurement of the cross terms would 
require a prior knowledge of the charge fluctuations for a single 
impedance with a high degree of accuracy. We argue that this is not the 
case, provided that two measurements with the same setup but with different
wiring can be achieved. One measures the charge fluctuations 
$\langle x^2(0)\rangle_{2+}$ with the geometry 
of Fig.~\ref{fig:setup}a for the inductances in series (Fig.~\ref{fig:setup}c for 
the parallel case), and subsequently one can switch the wiring and measure 
such fluctuations $\langle x^2(0)\rangle _{2-}$ with the circuit of 
Fig.~\ref{fig:setup}b (Fig.~\ref{fig:setup}d). In each case (series or parallel 
setup) by subtracting the two signals: 
  \be
 \langle x^2(0)\rangle _2=
\frac{1}{2}\big(\langle x^2(0)\rangle _{2+}-\langle x^2(0)\rangle _{2-}\big)
~,  \ee
one isolates the contribution of cross-correlations, which is 
proportional to $\alpha_1\alpha_2$. This combination of charge fluctuations, which in turn 
depends on current 
cross-correlators, will be referred from now on as the measured cross 
correlations.

\section{Measured cross-correlations and non-symmetrized noise}

In order to proceed, the charge and the momentum are now written in terms of the oscillator variables 
of the LC circuit: 
\be
x(t)=\sqrt{\frac{\hbar}{2M\Omega}}(a e^{-i\Omega t}
+a^{\dag}e^{i\Omega t})~,
\ee
\be
p(t)=i\sqrt{\frac{\hbar M\Omega}{2}}(a^{\dag}e^{i\Omega t}-
a e^{-i\Omega t})~,
\ee
where $a$ is the destruction operator which satisfies $\langle a^\dag a \rangle=N(\Omega)$. 
The first commutator in Eq.~(\ref{two_commutators}) becomes: 
\be
[x^2(0),p(t_1) I_i(t_1)]=
    \frac{2i\hbar}{M\Omega}\Big(\frac{\hbar M\Omega}{2}\Big)^{1/2}\cos({\Omega t_1})(a+a^\dag)I_i (t_1) ~,
\ee
 and the average of the two interlocked correlators of Eq.~(\ref{two_commutators}) 
 reads:
   \be
    \lefteqn{\Big\langle\Big[[x^2(0),p(t_1) I_i(t_1) ],p(t_2) I_j(t_2) \Big]
\Big\rangle
    =-\hbar^{2}\cos({\Omega t_1})}\label{imbricatedcommutator}\nonumber\\
    &\times &\Big\{\langle I_i(t_1) I_j(t_2)\rangle\big[
      (N(\Omega)+1)e^{i\Omega t_2}-N(\Omega)e^{-i\Omega t_2}\big]\nonumber\\
    && -\langle I_j(t_2) I_i(t_1)\rangle\big[N(\Omega)e^{i\Omega t_2}
      -(N(\Omega)+1)e^{-i\Omega t_2}\big]\Big\}~,
   \ee
with $i,j=1,2$ ($i\neq j$). Substituting this 
commutator in the expression of charge fluctuations, 
four correlators of current derivatives appear in 
the result. Translationnal invariance motivates the change of variables: 
$\{t_1,t_2\}\to \{t=t_1-t_2,T=t_1+t_2\}$. 
The charge fluctuations become:
 \be\label{fluct}
\lefteqn{\langle x^2(0) \rangle = \frac{\alpha_1\alpha_2}{(4M)^2}
    \int_{-\infty}^{+\infty}dt\int_{-\infty}^{0}dTe^{\eta T} }\nonumber\\
    &\times &\Big\{(e^{i\Omega t}+e^{-i\Omega T sign(t)}) \nonumber\\
    & &\times[(N(\Omega)+1)(\langle I_{2}(0) I_{1}(t)\rangle
      +\langle I_{1}(0) I_{2}(t)\rangle)
\label{ttoinfinity}\nonumber\\
    & &  -N(\Omega)(\langle I_{1}(t) I_{2}(0)\rangle
      +\langle I_{2}(t) I_{1}(0)\rangle)]\Big\} ~.
  \ee 
In a manner similar to Refs.~\onlinecite{lesovik_loosen} and \onlinecite{gavish}, the following
non-symmetrized current correlators are introduced in Fourier space:
  \be
    S^+_{ij}(\omega)&=&\int \frac{dt}{2\pi} e^{i\omega t}
    \langle I_i(0)I_j(t)\rangle~,\label{C+}
   \\
    S^-_{ij}(\omega)&=&\int \frac{dt}{2\pi} e^{i\omega t}
    \langle I_i(t)I_j(0)\rangle~.\label{C-}
  \ee
Translation invariance relates these two correlators: 
$S^-_{ij}(\omega)=S^+_{ij}(-\omega)$, and furthermore
$\langle I_i(\omega_1)I_j(\omega_2)\rangle=\delta(\omega_1-\omega_2)S^+_{ij}(\omega_2)$, 
where $I_i(\omega)$ is the Fourier transform of $I_i(t)$.
Note that only in the case of autocorrelation: $S^+_{ii}(\omega)$ 
($S^-_{ii}(\omega))$ can be identified as an emission (absorption) rate from the 
mesoscopic device at positive frequencies.
At this point, both integrals in Eq.~(\ref{fluct}) can be performed.
The integration over $t$ gives two contributions:
one is a delta function, and the other gives products of principal 
parts, which cancel out. 
The charge fluctuations take the final form:
\be
    \langle x^2(0)\rangle &=&\frac{\pi\alpha_1\alpha_2}{2\eta(2M)^2}
            \left[(N(\Omega)+1)(S^+_{12}(\Omega)+S^+_{21}(\Omega))
              \right.\nonumber\\
& &\left. -N(\Omega)(S^-_{12}(\Omega)+S^-_{21}(\Omega))\right]~.
\label{chargefluctu}
  \ee  
To show that $\langle x^2(0)\rangle$ is real, 
we use the properties of the noise correlators:
$[S^\pm_{ij}(\Omega)]^*=S^\pm_{ji}(\Omega)$,
and the measured charge fluctuation reads:
\begin{eqnarray}
   \langle x^2(0)\rangle &=&\frac{\pi\alpha_1\alpha_2}{\eta(2M)^2}
            Re\left[(N(\Omega)+1)S^+_{12}(\Omega)
              \right.\nonumber\\
& &\left. -N(\Omega)S^-_{12}(\Omega)\right]~.
\end{eqnarray}
This is a central result of this paper, which is illustrated
below for a specific mesoscopic circuit.  The LC measurement setup effectively measures 
the real part of the noise cross correlator. 
A similar result was mentioned in Ref.~\onlinecite{lebedev_collapse}, although without 
justification.
Note that the final result still depends on the adiabatic 
coupling parameter.
In order to eliminate the dependence on $\eta$, the calculation can be generalized 
to a system with a distribution of oscillators, peaked around $\Omega$.
This is discussed in the appendix.
The fact that the LC circuit has a finite line shape, has of course a 
physical origin: the LC circuit contains dissipative elements (due to both the finite 
conductivity of the wires and to the electronic environment which surround 
the circuit), but for simplicity we do not consider the detailed mechanism 
for dissipation here.

\section{Measurement with two LC circuits}

\begin{figure}[h]
	\begin{center}
	\includegraphics[width=5 cm]{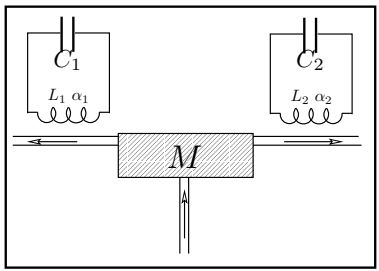}
	\end{center}

\caption{Schematic description of the noise cross-correlation setup. $M$ is 
the mesoscopic circuit to be measured, $C_1$ and $C_2$ are the two capacitors and there are 
two inductors with coupling constants $\alpha_1$ and $\alpha_2$
to the mesoscopic circuit. 
}
\label{fig:setup_2}
\end{figure}

Next, we examine the situation for using 
two separate circuits, with capacitors ($C_1$ and $C_2$) and two inductances ($L_1$ and $L_2$)
with inductive coupling constants
 $\alpha_1$ and $\alpha_2$. The two LC circuits  
are placed next to the two outgoing arms
of the three terminal mesoscopic device. 
The two capacitors is charging by the two inductances (Fig.~\ref{fig:setup_2}). 

The charges on each capacitor satisfy equations of motion similar to 
the previous calculation with a single circuit, 
%
where the ``mass'' $M_{1(2)}=L_{1(2)}$.
The characteristic frequency of each $LC$ circuit is $\Omega_{1(2)}=\sqrt{1/C_{1(2)}M_{1(2)}}$.


The second power of the capacitor two-charge correlator
reads, in the interaction representation: 
 \be 
\langle x_1(0)x_2(0)\rangle=Tr[e^{-\beta (H_{0,1}+H_{0,2})}U^{-1}(0)x_1(0)x_2(0)U(0)]~.
 \ee
Note that unlike in the previous section, we do not provide a scenario here for 
finding an observable which corresponds to the measurement of the product of
the two charges. 
We calculate perturbatively the $n$-th power of the charge,  
expanding the evolution term ($U(0)$) in powers of $H_{int,1(2)}$.
Considering the average charge and the average charge square, one obtains:
\be
\langle x_{1(2)}(0)\rangle&=&
\langle x_{1(2)}(0)\rangle_1+
\langle x_{1(2)}(0)\rangle_3+...~,
\\
\langle  x_1(0)x_2(0)\rangle&=&
\langle  x_1(0)x_2(0)\rangle_0+
\langle x_1(0)x_2(0)\rangle_2+...~,
\ee
where the different orders in $H_{int,1(2)}$
are linked to the higher cumulants of the current:
$\langle x_{1(2)}(0)\rangle_1$ contains information 
about the average current, the product $\langle x_1(0)x_2(0)\rangle_2$
about the current fluctuations, 
and $\langle x_{1(2)}(0)\rangle_3$ about the third moment, 
and so on. 

The zero order contribution of the charge fluctuation $\langle x_1(0)x_2(0)\rangle_0$ becomes zero
  . The first non-vanishing term, which 
depends on products of current operators is:
 \be
    \langle x_1(0)x_2(0) \rangle_{2}=&&
    \Big\langle
\frac{1}{2}
		\int_{-\infty}^{0}dt_1
                \int_{-\infty}^{t_1}dt_2        \nonumber\\
                 &\times    & \left[
                        [x_1(0)x_2(0),\frac{-i}{\hbar M_1} p_1(t_1)\alpha_1 I_1(t_1)+ \frac{-i}{\hbar M_2}p_2(t_1)\alpha_2 I_2(t_1)],\right.\nonumber\\
                &&\left.  \frac{-i}{\hbar M_1} p_1(t_2)\alpha_1 I_1(t_2)+\frac{-i}{\hbar M_2}p_2(t_2)\alpha_2 I_2(t_2)     
		\right]
                     \Big\rangle\nonumber\\
		&- & \frac{1}{2}\Big(\frac{-i}{\hbar}\Big)^2
                \int_{-\infty}^{0}dt \Big\langle\Big( \frac{\alpha_1 I_1(t)}{M_1}\Big)^2+\Big( \frac{\alpha_2 I_2(t)}{M_2}\Big)^2\Big\rangle\langle x_1(0)x_2(0)\rangle \nonumber\\
	      && -\langle x_1(0)x_2(0)\rangle \Big\langle\Big( \frac{\alpha_1 I_1(t)}{M_1}\Big)^2+\Big( \frac{\alpha_2 I_2(t)}{M_2}\Big)^2\Big\rangle ~.
  \label{two_commutators_twocapa}
  \ee
The calculation of the charge fluctuations gives four terms: two 
autocorrelation terms which correspond to the fluctuations due to a single 
inductor (in terms of $\alpha_1^2$ or 
$\alpha_2^2$), and two terms associated with the correlation 
between the two inductors, which are proportional to $\alpha_1\alpha_2$.
In the two autocorrelation terms, we calculate the average of an odd number of creation and destruction operators for the each circuit. The autocorrelation is zero. 
The combination of current 
cross-correlators will be referred from now on as the measured cross-correlations. 

Choosing a particular case where the two $LC$ circuits have the same characteristic frequencies ($\Omega_1=\Omega_2=\Omega$), the charge fluctuations 
correspond to the result obtained in the presence of a single capacitor with two inductances (see Eq.~\ref{chargefluctu}). 
In practice, it is challenging to build two LC circuits with exactly the same characteristic frequency. The result 
of Eq. (\ref{chargefluctu}) for $\Omega_1=\Omega_2$ has also appeared in Ref. \cite{gavish_phd}.  
For the case of different frequencies ($\Omega_1\neq\Omega_2$), The charge fluctuations have a general form:
   
\begin{eqnarray}
\lefteqn{\langle x_1(0)x_2(0) \rangle=-\frac{\alpha_1\alpha_2}{2}\int_{0}^{+\infty}dt  \Big\{}\nonumber\\
&\times&(\frac{e^{i t (\Omega_1-\Omega_2)/2}}{\eta+i\frac{\Omega_1+\Omega_2}{2}}
+\frac{e^{-i t (\Omega_2+\Omega_1)/2}}{\eta+i\frac{-\Omega_1+\Omega_2}{2}})
[(N(\Omega_2)+1)\langle I_{1}(t) I_{2}(0)\rangle-N(\Omega_2)\langle I_{2}(0) I_{1}(t)\rangle]\nonumber\\
&+&(\frac{e^{i t (\Omega_2+\Omega_1)/2}}{\eta+i\frac{\Omega_1-\Omega_2}{2}}
+\frac{e^{i t (\Omega_2-\Omega_1)/2}}{\eta-i\frac{\Omega_1+\Omega_2}{2}})
[N(\Omega_2)\langle I_{1}(t) I_{2}(0)\rangle-(N(\Omega_2)+1)\langle I_{2}(0) I_{1}(t)\rangle]\nonumber\\
&+&(\frac{e^{i t (\Omega_2-\Omega_1)/2}}{\eta+i\frac{\Omega_1+\Omega_2}{2}}
+\frac{e^{-i t (\Omega_1+\Omega_2)/2}}{\eta+i\frac{\Omega_1-\Omega_2}{2}})
[(N(\Omega_1)+1)\langle I_{2}(t) I_{1}(0)\rangle-N(\Omega_1)\langle I_{1}(0) I_{2}(t)\rangle]\nonumber\\
&+&(\frac{e^{i t (\Omega_1+\Omega_2)/2}}{\eta+i\frac{-\Omega_1+\Omega_2}{2}}
+\frac{e^{i t (\Omega_1-\Omega_2)/2}}{\eta+i\frac{-\Omega_1-\Omega_2}{2}})
[N(\Omega_1)\langle I_{2}(t) I_{1}(0)\rangle-(N(\Omega_1)+1)\langle I_{1}(0) I_{2}(t)\rangle]\Big\}~,
\end{eqnarray}
which result in delta function contributions, as well as (unwanted) principal parts: 
strictly speaking the charge fluctuations do not simplify.
However, a practical measurement always involves some bandwidth 
averaging, which could possibly lead to a result close to that of 
the case  $\Omega_1=\Omega_2$.

Furthermore, the use of two separate capacitors should also consider the multiplication process 
of the two charge operators. Multiplication of two signals involves adding and subtracting
the two signals, then squaring and subtracting. In these steps of the measurement, the noise from the 
amplifiers will be detrimental in the same manner as in the case of a single LC circuit.

\section{Application to a three terminal normal conductor}

\begin{figure}[h]
\center
\includegraphics[width=5cm]{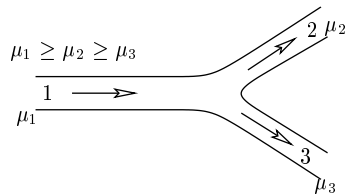}
\caption{A system with three terminals (Y junction) with chemical potentials  $\mu_1$, $\mu_2$ and $\mu_3$ with  $\mu_1\geq\mu_2\geq\mu_3$.}
\label{threeterm}
\end{figure}

Next, we consider noise measurement with the single capacitor and two inductances circuit initially proposed.
It is tested on a system 
of three terminals (Fig \ref{threeterm}), a so called ``Y junction'', 
electrons are injected from terminal $1$, which has a higher chemical 
potential than terminals $2$ and $3$. The noise cross--correlations 
are measured. This corresponds to the setup where a fermionic analog of Hanbury--Brown 
and Twiss type experiments\cite{HBT} was proposed\cite{martin_landauer} and measured\cite{henny_oliver}. Without loss of generality, one
considers that the three bias voltages, $\mu_{ij}=\mu_i-\mu_j$, ($i,j=1,2,3$),
are chosen such that $\mu_{13}>\mu_{12},\mu_{23}$.

Using scattering theory, one 
can readily obtain the general expression for the non-symmetrized finite frequency 
noise\cite{martin_houches}:
\be
S^+_{\alpha\beta}(\omega)&=&\frac{e^2}{2\pi \hbar}\sum_{\gamma \delta}\int dE\
 (\delta_{\alpha\gamma}\delta_{\alpha\delta}-s_{\alpha\gamma}^\dag(E)s_{\alpha\delta}(E-\hbar\omega))\nonumber\\
\qquad & &\qquad\qquad  \times (\delta_{\beta\delta}\delta_{\beta\gamma}-s_{\beta\delta}^\dag(E)s_{\beta\gamma}(E-\hbar\omega))
\nonumber\\
\qquad & & \qquad\qquad \times f_\gamma(E)(1-f_\delta(E-\hbar\omega))~,\label{general_noise_formula}
\ee 
where Greek letters represent the terminals and $s_{\alpha,\beta}$ is the scattering amplitude for electrons
incoming from $\beta$ and ending in $\alpha$. $f_\gamma(E)$ is the Fermi-Dirac distribution function associated with terminal $\gamma$ 
whose chemical potential is $\mu_\gamma$. In what follows we assume that the temperature
is much smaller than the applied biases.
One also neglects the energy dependence of the 
scattering matrix over the energy ranges specified by the voltages biases $\mu_{ij}$. 
Furthermore, Eq.~(\ref{general_noise_formula}) neglects $\pm 2k_F$ oscillating terms 
in the noise fluctuations spectrum: this assumes that the region over which current 
is measured is much larger than the Fermi wave length, $l\gg \lambda_F$.

For $\mu_{23}<\mu_{12}$ and at negative frequencies, cross-correlations between terminals $2$ and $3$ yield:
\be \label{ns1}
S^+_{23}(\omega)&=&S^+_{32}(\omega)=
-\frac{e^2}{2\pi }\omega(T_{21}T_{13}-(2-R_2-R_3)T_{23}  )\nonumber\\&&
-\frac{e^2}{2\pi \hbar}
\left\{
\begin{array}{ll}
(- \hbar\omega)(2T_{13}R_3   +2T_{12}R_2+2T_{23}R_2)                           &  \textrm{if }\hbar\omega<-\mu_{13}\\
T_{13}R_3   (\mu_{13}-\hbar\omega)     - \hbar\omega(2T_{12}R_2+2T_{23}R_2)  &  \textrm{if }-\mu_{13}<\hbar\omega<-\mu_{12}\\
T_{13}R_3   (\mu_{13}-\hbar\omega)+T_{12}R_2(\mu_{12}-\hbar\omega)- \hbar\omega(2T_{23}R_2)
          &  \textrm{if }-\mu_{12}<\hbar\omega<-\mu_{23}\\
T_{23}R_2(\mu_{23}-\hbar\omega)+T_{13}R_3   (\mu_{13}-\hbar\omega)+T_{12}R_2(\mu_{12}-\hbar\omega)  
 &  \textrm{if }-\mu_{23}<\hbar\omega<0
\end{array}
\right. \nonumber\\
\ee
while for positive frequencies:
\be \label{ns2}
S^+_{23}(\omega)&=&S^+_{32}(\omega)=
-\frac{e^2}{2\pi \hbar}\left\{
\begin{array}{ll}
T_{23}R_2(\mu_{23}-\hbar\omega)+T_{13}R_3   (\mu_{13}-\hbar\omega)+T_{12}R_2(\mu_{12}-\hbar\omega)    &  \textrm{if }0<\hbar\omega<\mu_{23}\\
T_{13}R_3   (\mu_{13}-\hbar\omega)+T_{12}R_2(\mu_{12}-\hbar\omega)      &  \textrm{if }\mu_{23}<\hbar\omega<\mu_{12}\\
T_{13}R_3   (\mu_{13}-\hbar\omega)             &  \textrm{if }\mu_{12}<\hbar\omega< \mu_{13}\\
0                   &  \textrm{if } \mu_{13}<\hbar\omega
\end{array}
\right.\nonumber\\
\ee

where  $R_\alpha=s_{\alpha,\alpha}^\dag s_{\alpha,\alpha}$ is the reflection probability from lead $\alpha$ and $T_{\alpha\beta}=s_{\alpha,\beta }^\dag s_{\alpha,\beta }=T_{\beta\alpha}$ is the transmission probability from $\alpha$ to $\beta$.

As expected, the frequency dependence is given by a set of continuous straight lines with 
singular derivatives. 
At zero frequency, the non-symmetrized cross-correlations are:
\be \label{ns}
S^+_{23}(\omega=0)&=&-\frac{e^2}{2\pi\hbar}
(T_{23}R_2\mu_{23}+T_{13}R_3\mu_{13}\nonumber\\
&~&~~~~
+T_{12}R_2\mu_{12})~.
\ee
The cross-correlations are negative regardless of bias voltage and 
transmission of the sample: one recovers the result of Ref.~\onlinecite{martin_landauer}.

On the other hand, the symmetrized finite frequency cross-correlations are defined by: 
\be
S_{23}^S(\omega)=\int d\omega e^{i\omega t}\langle \Delta I_2(t)\Delta I_3(0) + \Delta I_3(0)\Delta I_2(t)\rangle~,
\ee

with $\Delta I(t)=I(t)-\langle I(t) \rangle$. From Ref.~\onlinecite{blanter_buttiker}, when 
$\mu_{23}<\mu_{12}$ and at zero temperature, one obtains:
\be
S_{23}^S(\omega)&=&-\frac{e^2}{ 2\pi      }     |\omega|(T_{21}T_{13}-(2-R_2-R_3)T_{23})\nonumber\\&&
-\frac{e^2}{2 \pi \hbar}     \left\{
\begin{array}{ll}
+T_{23}R_2\mu_{23}+ T_{13}R_3   \mu_{13}+ T_{12}R_2\mu_{12}  &  \textrm{if }|\hbar\omega|<\mu_{23}\\
  \hbar|\omega|( T_{23}R_2)+ T_{13}R_3   \mu_{13}+ T_{12}R_2\mu_{12}           &  \textrm{if }\mu_{23}<|\hbar\omega|<\mu_{12}\\
 \hbar|\omega|( T_{12}R_2+T_{23}R_2)+ T_{13}R_3   \mu_{13}                    &  \textrm{if }\mu_{12}<|\hbar\omega|< \mu_{13}\\
 \hbar|\omega|( T_{13}R_3   +T_{12}R_2+T_{23}R_2      )                       &  \textrm{if }\mu_{13}<|\hbar\omega|
\end{array}
\right. 
\ee

\begin{figure}[h]
\center
\includegraphics[width=8cm]{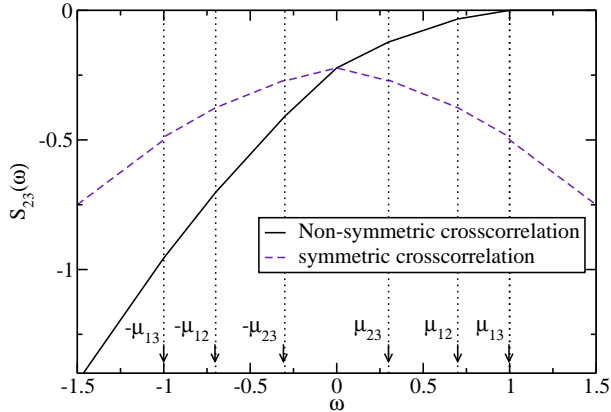}
\caption{Comparison between non-symmetrized and symmetrized cross-correlations 
as a function of frequency for $\mu_{12}>\mu_{23}$, in units of $e^2\mu_{13}/2\pi\hbar$. 
Singularities occur at $\hbar\omega=$ $0$, $\pm\mu_{23}$, $\pm\mu_{12}$, $\pm\mu_{13}$. The curves are plotted for 
$\mu_{12}=0.7\mu_{13}$ and 
$\mu_{23}=0.3\mu_{13}$.}
\label{symnsym}
\end{figure}

The frequency dependence of $S_{23}^S(\omega)$ is symmetric in $\omega$. The non-symmetrized cross-correlations, given by Eqs.~(\ref{ns1}) and (\ref{ns2}), coincides with the symmetrized cross-correlations at $\omega=0$, as illustrated in Fig.~\ref{symnsym}. The non-symmetrized cross-correlations behavior is quite different from the 
symmetrized cross-correlations, although the locations of their singularities
are the same. The symmetrized and non-symmetrized cross-correlations are both negative. 
The non-symmetrized cross-correlations are monotonously increasing and equal to zero for $\hbar\omega>\mu_{13}$.

We now consider the measured cross-correlations, given by Eq.~(\ref{chargefluctu}), as a function of 
temperature of the measurement circuit. Because of the symmetry of the 
transmission probabilities, 
we have in this case $S^\pm_{23}(\omega)=S^\pm_{32}(\omega)$
 \cite{lesovik_loosen,gavish}. 
With the above symmetry consideration, the charge fluctuations
resemble the formula for auto-correlation noise:
\be
\langle x^2(0)\rangle=\frac{\alpha_1\alpha_2\pi}{\eta(2M)^2}[
S^+_{23}(\omega)+\Delta S_{23}(\omega)N(\omega)]~.
\ee
Note that it is the difference of the two non--symmetrized correlators
$\Delta S_{23}(\omega)=S^+_{23}(\omega)-S^+_{23}(-\omega)$ which
is multiplied by the Bose distribution $N(\omega)$:
\be
\Delta S_{23}(\omega)&=&-\frac{e^2}{2\pi}\omega \Big(T_{21}T_{13}-(2-R_2-R_3)T_{23}\nonumber
\\
&&-2T_{13}R_3   -2T_{12}R_2-2T_{23}R_2\Big)~,
 \ee
i.e., it is linear with frequency with a positive slope, and does not have any singularities.

\begin{figure}[h]
\center
\vspace{0.8cm}
\includegraphics[width=8cm]{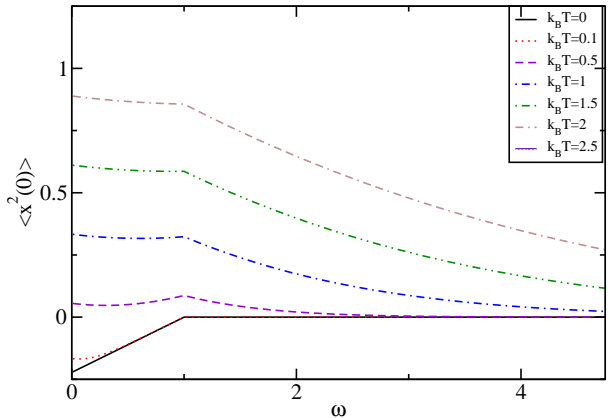}
\caption{Measured charge fluctuations as a function of frequency for different 
temperatures $k_BT$, measured in units of $\mu_{13}$. 
The frequency and the biases are in units of $\mu_{13}$.}
\label{crossHT1}
\end{figure}

Taking $\mu_{2}=\mu_{3}$ for simplicity, the measured cross-correlations are
plotted in Fig.~\ref{crossHT1}.

For temperatures such that $k_BT\gtrsim\mu_{13}$, 
the Bose distribution is large, $N(\omega)\sim k_BT/\hbar\omega\gg 1$. $\Delta S_{23}(\omega)N(\omega)$ 
is thus larger than $|S^+_{23}(\omega)|$. The effect of the increasing temperature is to change 
the sign of the charge fluctuations which becomes positive. 
This may seem change given the fact that one is computing the fluctuations of charge
at the plates of a capacitor: recall that our measurement implicitly assumes
two experiments with different wiring, whose results are subtracted one 
from another. The main message of Fig.~\ref{crossHT1} is that the sign of the 
 measured correlation can be misleading if the temperature of the measuring device
 is too large: one can observe positive cross-correlations in a normal fermionic 
 fork although this is a system where anti-bunching is expected.  
 For $\omega\geq\mu_{13}$, the measured noise equals $\Delta S_{23}(\omega)N(\omega)$, 
 while for $\omega<\mu_{13}$,  $S^+_{23}(\omega)$ lowers $\Delta S_{23}(\omega)N(\omega)$. 
 At all temperatures, the measured charge fluctuations have a clear singularity 
 at $\hbar\omega=\mu_{13}$: we are not taking into account the thermal effects
 in the mesoscopic circuit itself. 

 \begin{figure}[h]
 \center
\vspace{0.8cm}
     \includegraphics[width=8cm]{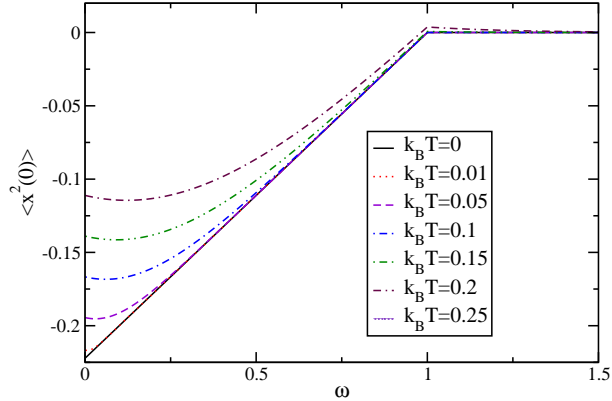}
 \caption{Same as Fig. \ref{crossHT1}, for a smaller frequency range and at lower temperatures
 (in units of $\mu_{13}$).}
 \label{crossBT1}
 \end{figure} 

 At low temperature, $k_BT \ll \mu_{13}$,  and low frequency, $\omega \ll \mu_{13}$, $\Delta S_{23}(\omega)N(\omega)$ 
 is smaller than $|S^+_{23}(\omega)|$ and the measured signal remains negative 
 (see Fig.~\ref{crossBT1}). When the temperature goes to zero, the measured charge 
 fluctuations equal $S^+_{23}(\omega)$.



%

For completeness, the case where the voltage biases satisfy 
$\mu_{13}>\mu_{12}>\mu_{23}>0$ is discussed. The fact that terminals 2 and 3 have different chemical potential will lower the contribution of $S^+_{23}(\omega)$ 
to $\Delta S_{23}(\omega)N(\omega)$, and increase the amplitude of charge fluctuations. At low temperature, the charge fluctuations stay negatives and singularities are present at frequencies equal to $\mu_{23}$, $\mu_{12}$, $\mu_{13}$. When the temperature becomes larger than the bias voltages, the charge fluctuations become positive. However, when the temperature goes to zero, the charge fluctuations  become equal to $S^+_{23}(\omega)$. When the difference between the chemical potentials of terminals $2$ and $3$ increases the effect of temperature is more important at small temperature the amplitude is bigger regardless of the effect of $S^+_{23}(\omega)$ is less important for large temperatures. 

\begin{figure}
\center
\vspace{0.8cm}
    \includegraphics[width=8cm]{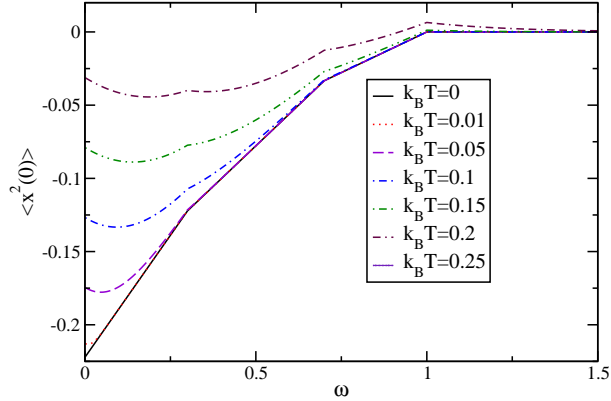}
\caption{Same as Fig. \ref{crossHT1}, for a smaller frequency range, and at lower temperatures.
$\mu_{12}=0.7\mu_{13}$, 
$\mu_{23}=0.3\mu_{13}$ have been chosen.
}
\label{crossBT2}
\end{figure}

\section{When are finite frequency cross-correlations noise useful ?}

When probing quantum non-locality with a source 
of electrons (for instance a S--wave superconductor
or any other source of electrons) 
with the help of Bell inequalities\cite{chtchelkatchev}, 
one is confronted 
with the fact that particle number correlators must be
converted into noise correlators. The particle number operator reads:
\be
N_\alpha(\tau)=\int_0^{\tau}I_\alpha(t')dt'=\langle N_\alpha(\tau)\rangle
+\delta N_\alpha(\tau)~,
\ee
where $I_\alpha(t)$ is the current operator in lead $\alpha$.
The irreducible particle number correlator is expressed in terms 
of the finite frequency shot noise power:
\be
\langle \delta N_\alpha(\tau)
\delta N_\beta(\tau)\rangle=(1/2\pi)\int_{-\infty}^{\infty}d\omega 
S_{\alpha,\beta}(\omega)4\sin^2(\omega\tau/2)/\omega^2~.
\ee 
It is therefore only in the limit of relatively ``large'' acquisition times
that the Bell inequality can be cast in terms of zero frequency 
correlations only\cite{chtchelkatchev,beenakker_entanglement}, 
thus the need in general to measure in general 
finite frequency noise correlations. This is implicit in the 
work of Ref.~\onlinecite{lebedev_entanglement}, where entanglement 
occurs with a normal source of electrons, provided 
that short time dynamics can be analyzed. 
Ref.~\onlinecite{chtchelkatchev} missed this subtlety 
concerning normal electron sources. 

Another situation of interest for finite frequency cross-correlations noise 
deals with the detection of anomalous (non-integer charges) in carbon nanotube. 
A Hanbury--Brown and Twiss experiment has been proposed where an STM tip injects electron
in the bulk of the nanotube and cross-correlations noise are measured at the 
extremities of the nanotube\cite{crepieux_guyon_devillard_martin,lebedev_crepieux_martin}. In Ref.~\onlinecite{crepieux_guyon_devillard_martin}, the case of an infinite nanotube has been considered. Schottky-like relations for the zero-frequency Fourier transforms of the
auto-correlation noise and the cross-correlations noise were derived within 
the Tomonaga-Luttinger model:
\be
S_{auto}(\omega=0)=\frac{1+(K_{c+})^2}{2}e|\langle I(x)\rangle |~,
\ee
\be
S_{cross} (\omega=0) =\frac{1-(K_{c+})^2}{2}e|\langle I(x)\rangle |~,
\ee
where $\langle I(x)\rangle$ is the charge current through the nanotube and $K_{c+}$ is the 
Luttinger liquid interaction parameter. $S_{auto}$ is the term of auto-correlations 
and $S_{cross}$ is the cross-correlations term. However, in the presence of electrical contacts at the extremities of the nanotube, the zero-frequency Fourier transforms for noise and cross-correlations lose their $K_{c+}$-dependence: at order 2 in the perturbative expansion with the tunneling amplitude from the tip to the nanotube, they reduce to $S_{auto}(\omega=0)=e|\langle I(x)\rangle |$ and $S_{cross}(\omega=0)=0$. It has been shown in Ref.~\cite{lebedev_crepieux_martin}, that one has to consider finite frequency Fourier transform in order to recovered non zero cross-correlations and coulomb interactions effects in such a system.


\section{Conclusion}

In summary, we have shown that the cross-correlations noise of a mesoscopic circuit can be
measured by coupling the latter to a resonant circuit which is composed of two inductances and a
capacitor. Each inductance is attached to the arms where correlations are measured.
As in Ref. \cite{lesovik_loosen}, the proposed noise measurement is made by monitoring
the charge fluctuations on the capacitor. Two distinct measurements, with different
wiring of the circuit, are necessary to isolate the cross-correlations noise.

Granted, we have not described the experimental apparatus which is needed to convert the 
signal into classical information, i.e. the amplification scheme: 
this goes beyond the scope of the present paper, 
which goal is to convey that information about cross-correlations noise can be 
converted into a charge signal on a capacitor. 
A general approach for analyzing the quantum behavior of electronic circuitry
has been proposed \cite{yurke}, allowing to treat both nonlinear and 
dissipative elements, hence useful for signal conversion. The transition from 
mesoscopic to macroscopic quantum transport has been addressed in the context 
of scattering theory \cite{liu}. 
Note that in our analysis of the setup, the variable $\dot{x}$ denotes the current 
flowing in the capacitor's circuit. It is assumed to be a constant current in space. 
This description may be inappropriate 
at high frequencies, setting some upper limit to the 
frequencies we want to probe. 

Note that the quality of the diagnosis of cross-correlations
presented in this work depends, to a
large extent, on how the mesoscopic circuit is perturbed
when one switches from one wiring configuration to the other.
Indeed, it is necessary to minimize the changes in
configuration (the values of $\alpha_1$ and $\alpha_2$)
of the circuit which occurs between the two measurements.
Here one could assume ideally that the inductances are built ``on-chip'' with the mesoscopic
circuit, or with the wires connected to it, which
is a challenge in practice: small inductors working at high frequencies ($\omega=100GHz$)
 may be difficult to achieve. While it may be more realistic to couple
 the two circuits away from the mesoscopic device, on chip inductances 
 of small scale are nevertheless used in qbit circuitry \cite{duty}.
 Moreover, the wires of the measurement circuit are assumed to 
 have a low impedance
compared to that of the mesoscopic sample: within these working
conditions the change of wiring does not affect significantly
the inductive couplings, and the measurement will be reliable.

As a first necessary application of this measurement scheme, we considered
a three terminal normal metal conductor, which is known to exhibit negative noise
correlations at zero frequency. The symmetrized noise differs strongly
from the non-symmetrized noise and the measured noise.
For a mesoscopic circuit at zero temperature,
the non-symmetrized noise contains singularities at frequencies corresponding
to the chemical potential differences. However, when considering the measured
noise, care must be taken to work with a detector circuit whose temperature
is below these relevant biases. In this case the singularities in the derivative
can still be detected, and upon increasing slowly the temperature, the measured noise
deviates from the non-symmetrized noise.
This provides a condition for the observation of negative
noise correlations -- electron anti-bunching -- in such three terminal devices.   
Because of the above mentioned temperature effects of the device, the
amplifiers needed for signal conversion would need to be cooled down
in order to avoid such problems.

The present analysis can be extended to study the measured noise
correlations in other mesoscopic devices, in particular
three terminal structures where electrons are injected
in one lead from a superconductor\cite{torres_martin}.
Finite frequency noise correlations in this case contain
information of the time dynamics of the two electrons subject to
crossed-Andreev transport. 

\appendix
\section{}

We discuss the case where the resonant LC circuit has a finite line shape.
For a distribution of oscillators, the charge operator now takes the form: 
\be
x(t)=\sum_\omega x_\omega(t)=\sum_\omega \sqrt{\hbar/2M\omega}
(a_\omega e^{-i\omega t} +a^\dag_\omega e^{i\omega t})~,
\ee 
where the interaction coupling, 
$
-\alpha I p/M$, becomes $-I\sum_\omega 
\alpha_\omega p_\omega/M$.

These expressions are substituted in the interlocked commutators 
Eq.~(\ref{two_commutators}), which now contains a sum over $4$ frequencies 
$\omega_1$, $\omega_2$,  $\omega_3$ and $\omega_4$.  
Because of time integrations, delta functions appear and give equalities between the 
frequencies. We are left with two summations:
\be
    \lefteqn{ 
      \Big\langle\Big[[x^2(0),p(t_1){I_i}(t_1)],
        p_(t_2){I_j}(t_2)\Big]\Big\rangle=} \nonumber\\
    & &{}
      \sum_{\omega_1,\omega_2}{\hbar^2}\cos(\omega_1 t_1)      
    \nonumber\\
    & &\times {}\Big[\langle {I}_i(t_1){I}_j(t_2)\rangle
      (N_{\omega_2}e^{-i\omega_2 t_2}-(N_{\omega_2}+1)e^{i\omega_2 t_2})
\nonumber\\
      & &{}-\langle {I}_j(t_2){I}_i(t_1)\rangle
      ((N_{\omega_2}+1)e^{-i\omega_2 t_2}-N_{\omega_2}e^{i\omega_2 t_2})
      \Big]~.
  \ee
We substitute this expression in the charge 
fluctuations and proceed with the change of variable, we make use of the 
time translational invariance: 
$t=t_1-t_2$ and $T=t_1+t_2$ and $\langle I_i(t)I_j(0)\rangle=\langle I_i(0)I_j(-t)
\rangle$.  
Making use of the definitions of Eqs.~(\ref{C+}) and (\ref{C-}),
the integral over $T$ leads to two contributions:
\be
K_1&=&\int_{-\infty}^0 dT e^{i(\omega_1+\omega_2)t/2}
e^{\eta T+i(\omega_1-\omega_2) sign(t)T/2}\nonumber\\
&=&\frac{ e^{i(\omega_1+\omega_2)t/2}}{\eta +i(\omega_1-\omega_2) sign(t)/2}~,\\
K_2&=&-\int_{-\infty}^0 dTe^{i(-\omega_1+\omega_2)t/2} 
e^{\eta T-i(\omega_1+\omega_2) sign(t) T/2}\nonumber\\
&=&-\frac{ e^{i(-\omega_1+\omega_2)t/2}}{\eta -i(\omega_1+\omega_2) sign(t)/2}~.
\ee
The line shape $L(\omega-\Omega)$, which is sharply peak around 
the resonant circuit frequency $\Omega$, is introduced when 
converting discrete sums over frequencies into integrals.
The integral over $t$ is performed subsequently.

The denominators of $K_1$ and $K_2$ yield a real part which are a 
principal part and an imaginary part which are a Dirac delta function. 
We obtain four contributions for the charge fluctuations:
\be
\langle x^2(0) \rangle= A_1+A_2+A_3+A_4~,
\ee

where 
 \be
    A_1   &=&4\pi^2\int d\omega L^2(\omega-\Omega)\frac{\alpha_1\alpha_2}{(2M)^2}\label{contributioA1}\nonumber\\ &&
    \times Re\left[(N_{\omega}+1)S^+_{12}(\omega) -N_{\omega}S^-_{12}(\omega)\right]~,
    \\
   A_2&=&-2\int d\omega_1 d\omega_2d\omega_3
    L(\omega_1-\Omega)L(\omega_2-\Omega)
   \frac{\alpha_1\alpha_2}{(2M)^2}
\nonumber\\
  && \times \mathcal{P}\Big(((\omega_1+\omega_2)/2-\omega_3)^{-1}\Big)
\mathcal{P}\Big(2/(\omega_1-\omega_2)\Big) \nonumber
\\
    & & \times  {} Re\left[(N_{\omega_2}+1)S^+_{12}(\omega_3) -N_{\omega_2}S^-_{12}(\omega_3) \right]~, \\
    A_3 &=&4\pi^2\int d\omega L  (\omega-\Omega)L (-\omega-\Omega)\frac{\alpha_1\alpha_2}{(2M)^2}\nonumber \\
    &&\times Re\left[(N_{\omega}+1)S^+_{12}(-\omega)
      -N_{\omega}S^-_{12}(-\omega)\right]~,
    \\
  A_4&=&-2\int d\omega_1 d\omega_2d\omega_3
    L(\omega_1-\Omega)L(\omega_2-\Omega)
     \frac{\alpha_1\alpha_2}{(2M)^2}\nonumber\\
 && \times  \mathcal{P}\Big(((\omega_1-\omega_2)/2-\omega_3)^{-1}\Big)
 \mathcal{P}\Big(2/(\omega_1+\omega_2)\Big)
 \nonumber
\\   
    & &\times {}Re \left[(N_{\omega_2}+1)S^+_{12}(\omega_3)
      -N_{\omega_2}S^-_{12}(\omega_3)\right]~,
  \ee
where the function $\mathcal{P}$ gives the principal part. The contribution which dominates is the contribution where the two line shape functions 
 $L(\omega-\Omega)$ are peaked at the same frequency, i.e., Eq (\ref{contributioA1}). 
 The quantity $\eta$ which appears in the single oscillator model thus corresponds
 physically to the width of the line shape, as expected.

\end{document}